\documentclass[runningheads]{llncs}
\usepackage{multirow}

\usepackage{subcaption}
\usepackage{xcolor}

\usepackage{graphicx}
\usepackage{hyperref}

\begin{document}
\title{Analyzing eyebrow region for morphed image detection}
%
%
\author{Abdullah Zafar\inst{1} \and
Christoph Busch\inst{1}}
\authorrunning{A. Zafar \& C. Busch}
%
\institute{Norwegian University of Science and Technology (NTNU), Gjøvik, Norway}
\newcommand{\red}[1]{#1}

\maketitle              

\begin{abstract}
Facial images in passports are designated as primary identifiers for the verification of travelers according to the International Civil Aviation Organization (ICAO) ~\cite{icao20159303}. Hence, it is important to ascertain the sanctity of the facial images stored in the electronic Machine-Readable Travel Document (eMRTD). With the introduction of automated border control (ABC) systems that rely on face recognition for the verification of travelers ~\cite{carlos2018facial}, it is even more crucial to have a system to ensure that the image stored in the eMRTD is free from any alteration that can hinder or abuse the normal working of a facial recognition system. One such attack against these systems is the face-morphing attack. Even though many techniques exist to detect morphed images, morphing algorithms are also improving to evade these detections. In this work, we analyze the eyebrow region for morphed image detection. The proposed method is based on analyzing the frequency content of the eyebrow region. The method was evaluated on two datasets that each consisted of morphed images created using two algorithms. The findings suggest that the proposed method can serve as a valuable tool in morphed image detection, and can be used in various applications where image authenticity is critical.

\end{abstract}

\section{Introduction} 
Face morphing is a real and live threat against the ABC systems which verify a person's identity by comparing the live image with the facial reference stored in the eMRTD ~\cite{zhang2018face}. Despite the simplicity of the solution of having the passport holder come to the center to take photographs, it is still not universally used due to financial reasons. Furthermore, many countries have adopted or are in the process of adopting web-based passport/visa applications for the ease of applicants where the user can upload a digital copy of the image to the web portal ~\cite{rajaface}. With technological advances, it is counter-intuitive for the general user to be asked to come to an office just to take a photograph. These things make the detection of a morphed image even more relevant in these times.

For simplicity, face morphing can be explained as an attack against face recognition systems where the images of two individuals are combined to create a morphed image that is used as a reference image. This reference image produces a positive match against the images that were used in creating the morphed image. One serious application of such an attack is explained in ~\cite{Ferrara2014TheMP} where face morphing enables an individual to travel on someone else's passport.

This paper will present a morphing attack detection (MAD) technique to detect morphed images during the enrollment phase. MAD methods can be divided into two categories i.e. single image MAD (S-MAD) and Differential MAD (D-MAD)~\cite{ndeh2021morphed}. S-MAD involves analyzing an image to determine whether the image is a morph or not. D-MAD involves analyzing the image and another trusted live source for detection. The approach presented in this paper is an S-MAD technique. However, the same technique can also be applied in a D-MAD scenario. S-MAD is more relevant for the passport application process because no trusted live source exists during the enrollment phase.

The approach is based on the analysis of the eyebrow region. The assumption is that the eyebrow region has a high-frequency content due to the presence of hairs. The idea is to analyze the possible reduction in this high-frequency information due to the smoothening effect that results from the averaging of two images in the creation of a morphed image. Furthermore, the eyebrow region is interesting because of its universality and importance in the performance of face recognition systems ~\cite{sadr2003role}. The rest of the paper is organized as: Section \ref{section:prior-work} will highlight some of the related work, Section \ref{section:methodology}  explains the methodology, Section \ref{section:results} presents the results obtained, and Section \ref{section:conclusion} concludes with final remarks and discussion.

\section{Prior Work}
\label{section:prior-work}
Morphing-based attacks against ABC systems were first identified by Ferrara et al. ~\cite{Ferrara2014TheMP} where they demonstrated the hypothetical scenario of a malicious actor who travels on his friend's passport by means of a face morphing attack. After that, the topic of morphed image detection piqued the interest of researchers resulting in a number of studies presenting different kinds of morphed detection techniques. In this section, past works are presented where texture descriptors are used for morphing attack detection.

Ramachandra et al. ~\cite{ramachandra2016detecting} in 2016 proposed the first single image based morphed attack detection system. It relied on texture descriptor differences in bonafide and morphed images. The algorithm worked by obtaining a micro-texture variation using Binarized Statistical Image Features (BSIF) and then making the decision using a linear support vector machine. The same detection technique was tested in ~\cite{scherhag2017vulnerability} against two databases of printed-scanned images. The reason for using print-scan images was to mimic the image quality in a visa application process as the visa application process in many countries requires submitting printed images that later get scanned to be saved in the system ~\cite{scherhag2018detecting}. The results showed that the detection performance of this technique dropped compared to the digital images. 

Spreeuwers et al. in ~\cite{spreeuwers2018towards} presented another MAD technique that was based on local binary patterns (LBD). Experiments were conducted on multiple databases and with different morphing algorithms to test the robustness of the proposed method. The results obtained were comparable to the BSIF-based method on one dataset, but the same performance could not be observed while testing on multiple datasets. 

The application of Fourier spectrum analysis on different facial \red{characteristics} was first suggested by Ndeh de Mbah in ~\cite{ndeh2021morphed}. This approach is based on analyzing the power density of six identified facial features. The decision of whether an image is morphed or bonafide is based on the total score obtained from the six classifiers. Experiments were conducted on two databases where the results varied greatly depending on the dataset used. However, the reasoning behind the difference in results was not addressed in the paper.

In this paper, we focus on analyzing the eyebrow region in the frequency domain to distinguish between a morphed and bonafide image. We use bonafide and morphed images from two different datasets containing ICAO-compliant printed scanned bonafide images and their morphs created using two morphing algorithms to test our approach. The proposed method is based on single-image detection that can also be used in a differential detection scenario.


\section{The Proposed Method}
\label{section:methodology}
The proposed method is a texture-based detection method where the smoothness property of the eyebrow region is studied to distinguish a morphed image from a bonafide image. The eyebrow region due to the presence of hairs is expected to have high-frequency content present in the Fourier domain. The study aims to find if this high-frequency content is lost due to the smoothening effect in a morphed image making it suitable to differentiate morphed images from bonafide images.

The experiments were based on developing a segmentation technique to crop the eyebrow region from a face image and then analyzing the segmented region in the frequency domain. The different steps of the experiment pipeline are described in the following subsections:

\subsection{Eyebrow region segmentation}
\begin{figure}
\begin{subfigure}{.49\textwidth}
  \centering
  \includegraphics[width=0.5\linewidth]{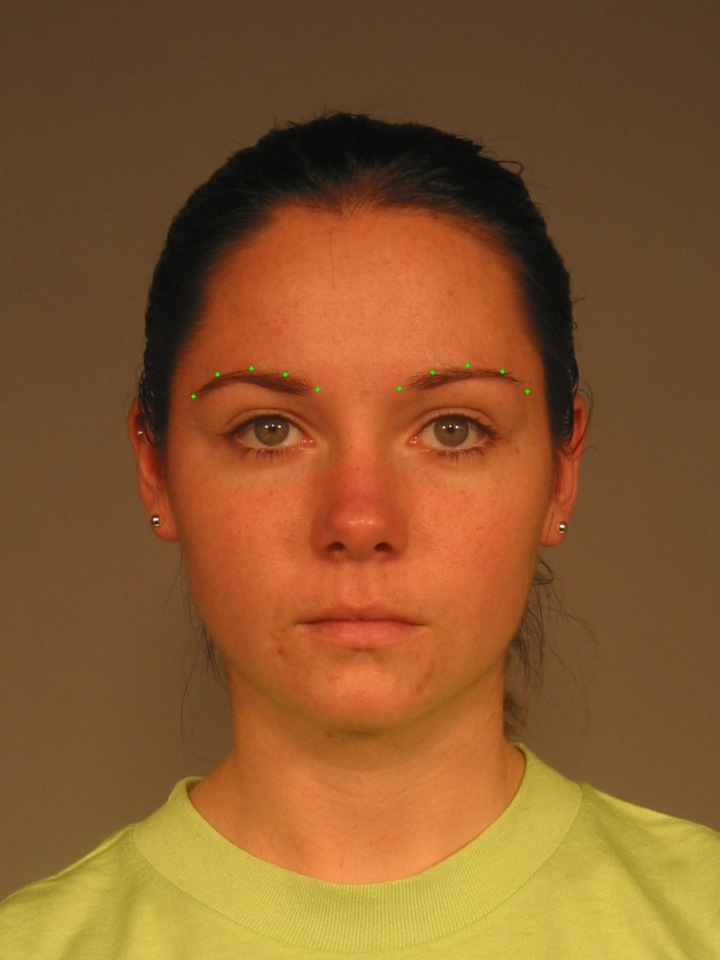}
  \caption{Sample image - Dlib landmarks}
  \label{fig:dlib-landmarks}
\end{subfigure}
\begin{subfigure}{.49\textwidth}
  \centering
    \raisebox{2.3\height}{\includegraphics[width=.8\linewidth]{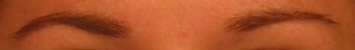}}

  \caption{Cropped eyebrow region}
  \label{fig:cropped-eyebrow-region}
\end{subfigure}
\caption{Cropping the eyebrow region}
\label{fig:eyebrow-region-cropping}
\end{figure}

For eyebrow region segmentation, we used Dlib's facial landmark detector to locate the eyebrow region in a face image. The pre-trained shape predictor provided by Dlib was used in this study ~\cite{dlibnetf3:online} which is based on the dataset from ~\cite{ibugres:online}. The Dlib shape predictor returns an array of length 68 containing coordinates of different facial features including the eyebrows. Eyebrows are marked by 10 array values with each eyebrow represented by 5 coordinates (Fig. \ref{fig:dlib-landmarks}). These eyebrow coordinates are then used to find the limits of the eyebrow region and crop a rectangle around the region as shown in Fig. \ref{fig:eyebrow-region-cropping}. This segmentation technique based on Dlib's landmark detector proved to be very effective for the ICAO-compliant images from the datasets.

\subsection{Pre-processing}
In this step, we prepare the cropped region for the frequency domain analysis. First, the image is converted to grayscale to be processed by the next stages. Converting the images to grayscale is important for the system to be used in a differential morph detection setting because some border control cameras only provide the trusted live capture image in grayscale ~\cite{scherhag2020face}. 

After converting the images to grayscale, the contrast of the cropped eyebrow region is increased to enhance the variations in the cropped image. \red{Contrast enhancement is done through black clipping and white clipping. It works by converting 1\% percentage of the darkest grey pixels to black (black clipping) and 5\% of the brightest grey pixels to white (white clipping), and then the rest of the gray pixels are scaled between the highest and the lowest values.}

\begin{figure}
\centering
\begin{subfigure}{.45\textwidth}
  \centering
  \includegraphics[width=0.8\linewidth]{images/bothEyebrows.jpg}
  \caption{Cropped eyebrow region}
\end{subfigure}
\begin{subfigure}{.45\textwidth}
  \centering
  \includegraphics[width=0.8\linewidth]{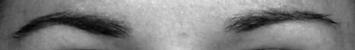}
  \caption{Enhanced grayscale image}
\end{subfigure}
\caption{Preprocessing the cropped image}
\label{fig:preproccesing}
\end{figure}

Since bonafide images are expected to have more variations, contrast enhancement is expected to further enhance these variations. Contrary to this, the morphed images because of their smoothened nature will be relatively less affected by this step. This phenomenon is also shown in the results in section \ref{section:contrast-results}. Fig. \ref{fig:preproccesing} shows the image after going through \red{the pre-processing step}.

\subsection{Fourier analysis}
As explained earlier, the idea is to differentiate morphed images from bonafide images by observing the smoothening of eyebrows in morphed images. In a sharp image, hairs in the eyebrows can be observed as edges separate from each other. These edges are represented by the high-frequency content in the frequency domain. 

\begin{figure}[ht]
\centering
\begin{subfigure}{.45\textwidth}
  \centering
  \includegraphics[width=\linewidth]{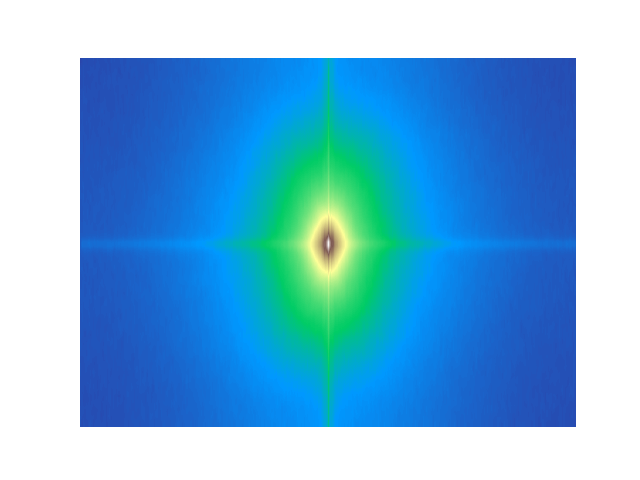}
  \caption{Bonafide}
\end{subfigure}
\begin{subfigure}{.45\textwidth}
  \centering
  \includegraphics[width=\linewidth]{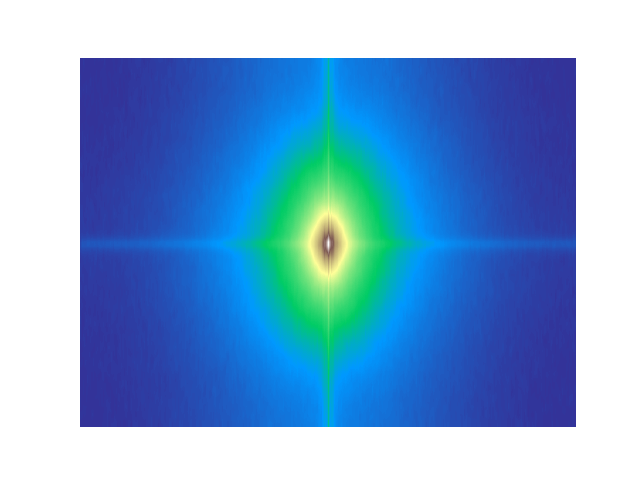}
  \caption{Morhped}
\end{subfigure}
\caption{Averaged DFT magnitude spectra of eyebrow regions of bonafide and morphed images}
\label{fig:dft-plots}
\end{figure}

Next, 2D Fourier transform of the preprocessed image is calculated to get the frequency representation of the image. Since the interest here is in the strength of the frequency content, only the magnitude of the Fourier transform is considered in the analysis. Fig. \ref{fig:dft-plots} shows the averaged DFT magnitude spectra of the eyebrow region of bonafide images and morphed (FaceFusion) images from the FRGC dataset. 

The plots are shifted to move the values associated with zero frequency to the middle \red{so that the} frequency increases as we move away from the origin. It can be observed from the DFT magnitude spectra that the outer circle for bonafide images is bigger than the morphed images indicating a wider spread out of high-frequency content in bonafide images. \red{In our approach, we exploit this difference in the frequency content to distinguish between morphed and bonafide images.}

\subsection{Calculating frequency content}
Once we have the Fourier spectrum, the next step is to establish a way of calculating the frequency content. We calculated the frequency content by taking the sum of the complete magnitude spectrum. The normalized sum calculated in our testing is expressed by the equation \ref{equation:sum-of-freq}.
\begin{equation}
sum = \frac{1}{MN} \sum_{n=1}^{N} \sum_{m=1}^{M} f(n,m)  
\label{equation:sum-of-freq}
\end{equation}
Here, f is an MxN array of the 2-dimensional DFT magnitude of the image, and M and N are the length and width of the cropped eyebrow region, respectively. The sum of coefficients is divided by the number of pixels in the cropped region for generalizing because the size of the cropped region can vary among different people and images of different resolutions.


\section{Experimental Results}
In the following subsections, the datasets used in the experimentation, evaluation metrics, the experiment setup, and results are presented.
\label{section:results}

\subsection{Datasets}
The morphed and bonafide images used in this experimentation are taken from ~\cite{scherhag2020face}. The bonafide images belong to two different datasets i.e. FERET ~\cite{phillips1998feret} and FRGCv2 ~\cite{phillips2005overview}. As described in ~\cite{scherhag2020face}, morphs are created by choosing the two subjects among the same dataset. In addition, the subjects are chosen based on their sex and whether they are wearing glasses. 

622 bonafide images from the FERET dataset and 1440 images from the FRGC dataset were used in this experiment. Both the bonafide and morph images are post-processed by passing through a print and scan pipeline to mimic the post-processing steps followed in a passport application process.

\begin{figure}
\begin{subfigure}{.24\textwidth}
  \centering
  \includegraphics[width=0.8\linewidth]{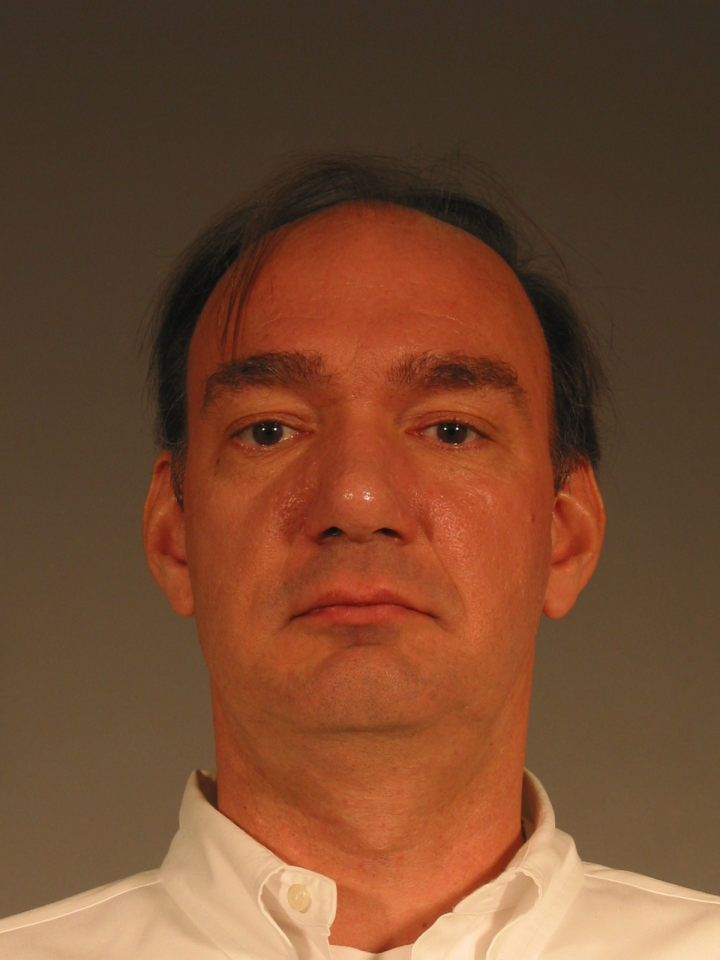}
  \caption{Subject 1}
\end{subfigure}
\begin{subfigure}{.24\textwidth}
  \centering
  \includegraphics[width=.8\linewidth]{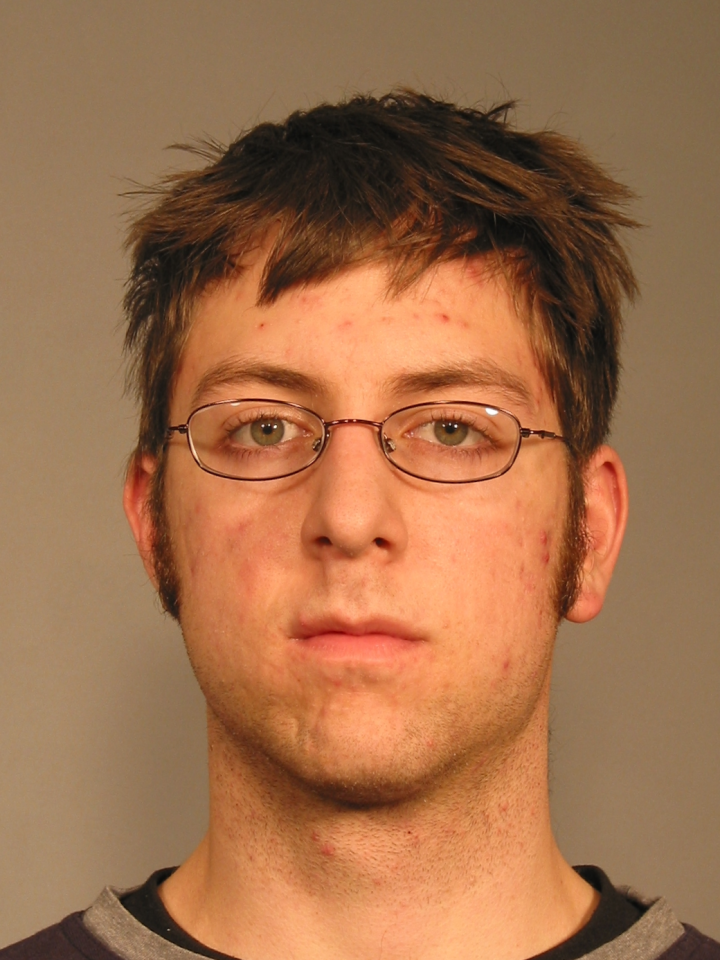}
  \caption{Subject 2}
\end{subfigure}
\begin{subfigure}{.24\textwidth}
  \centering
  \includegraphics[width=.8\linewidth]{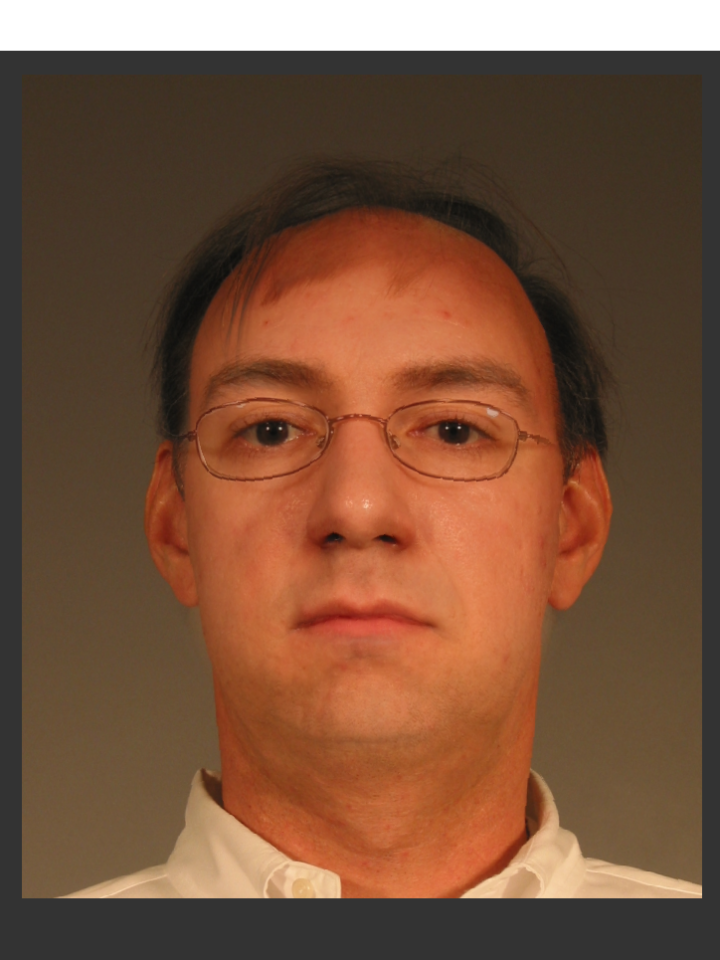}
  \caption{Facefusion}
\end{subfigure}
\begin{subfigure}{.24\textwidth}
  \centering
  \includegraphics[width=.8\linewidth]{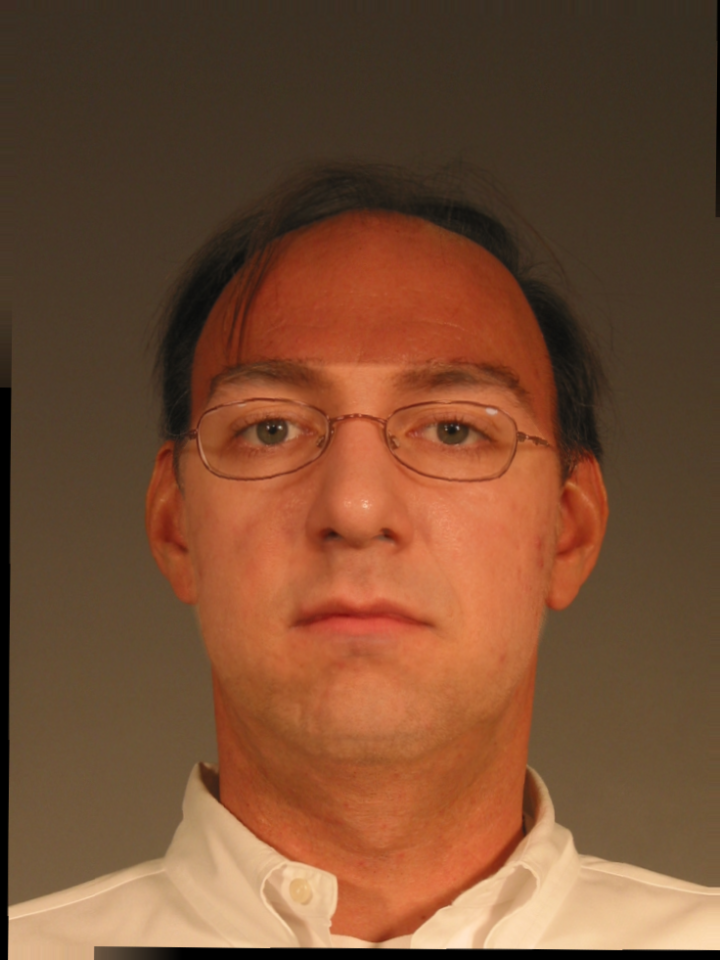}
  \caption{UBO Morpher}
\end{subfigure}
\caption{Sample images from FRGC subset and their morphs}
\label{fig:FRGC-samples}
\end{figure}

\begin{figure}
\begin{subfigure}{.24\textwidth}
  \centering
  \includegraphics[width=0.8\linewidth]{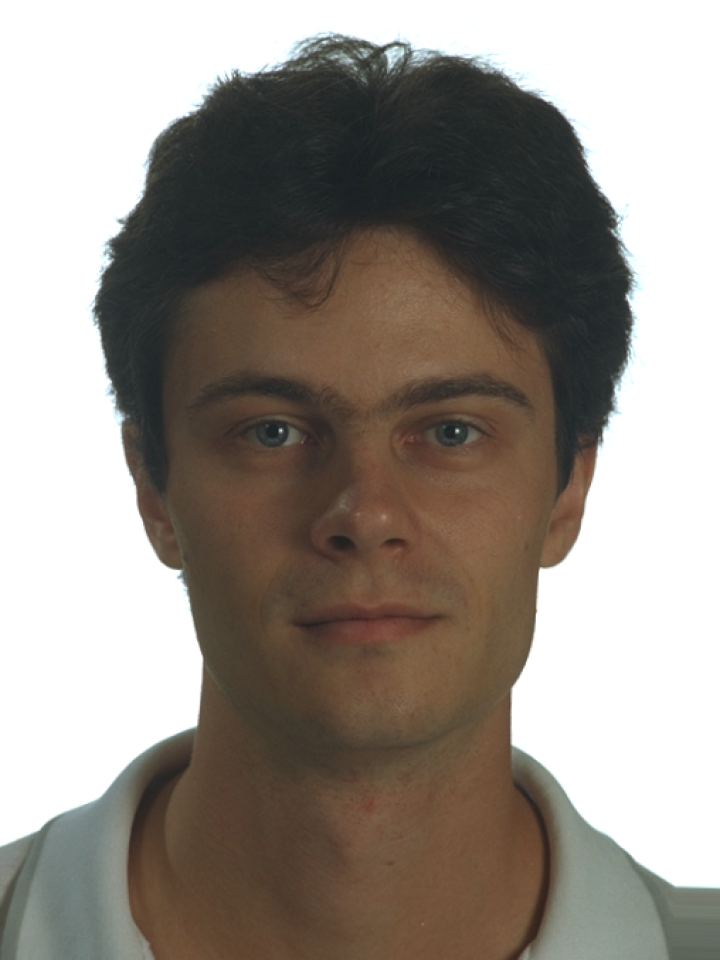}
  \caption{Subject 1}
\end{subfigure}
\begin{subfigure}{.24\textwidth}
  \centering
  \includegraphics[width=.8\linewidth]{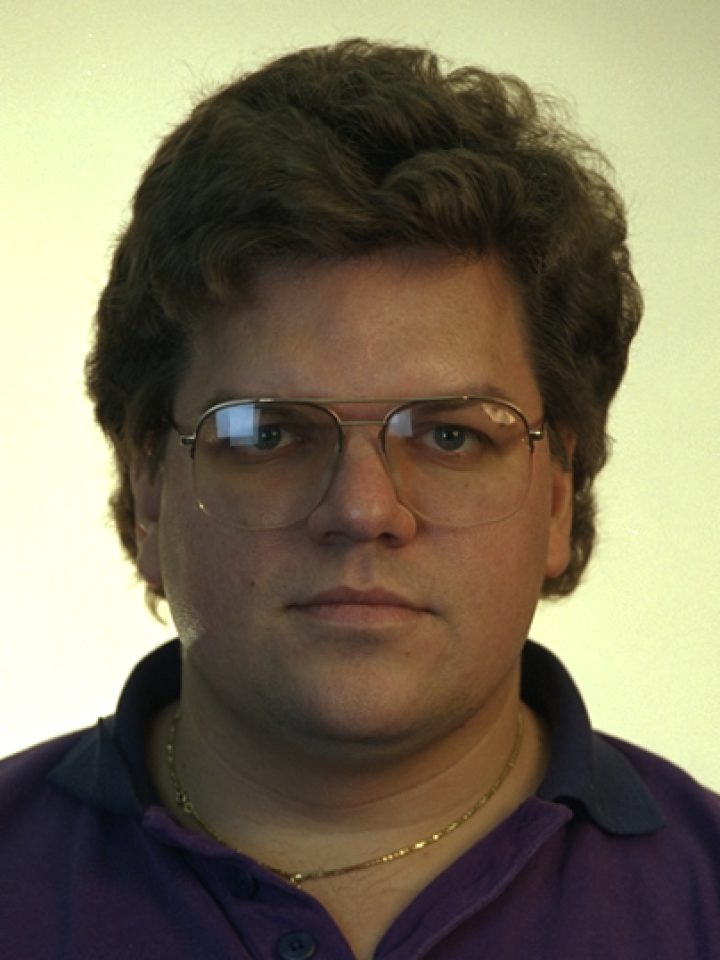}
  \caption{Subject 2}
\end{subfigure}
\begin{subfigure}{.24\textwidth}
  \centering
  \includegraphics[width=.8\linewidth]{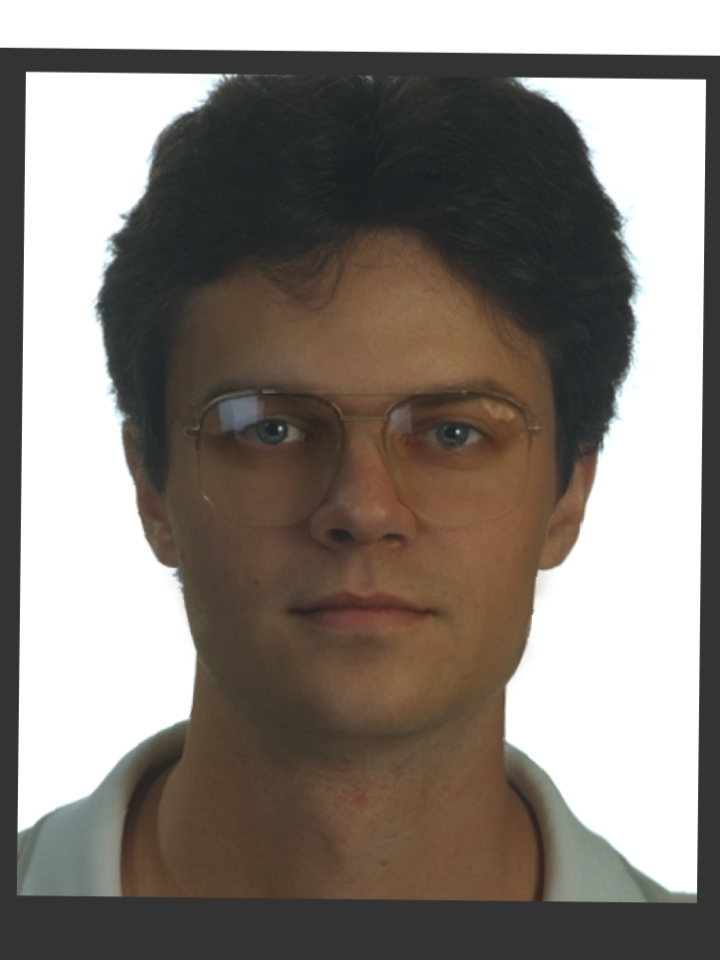}
  \caption{FaceFusion}
\end{subfigure}
\begin{subfigure}{.24\textwidth}
  \centering
  \includegraphics[width=.8\linewidth]{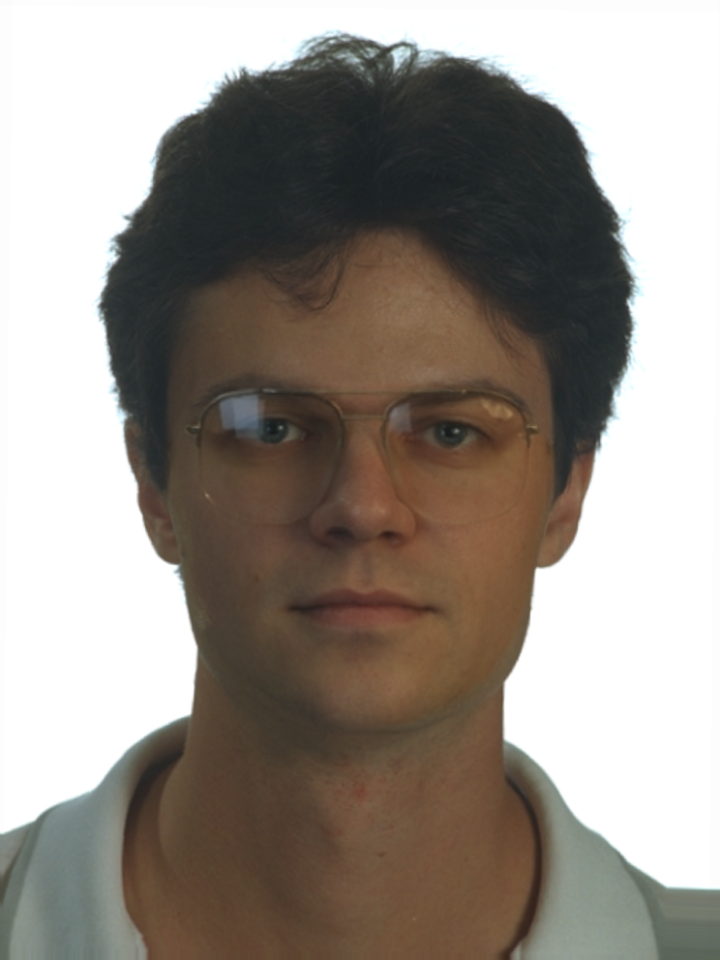}
  \caption{UBO Morpher}
\end{subfigure}
\caption{Sample images from FERET subset and their morphs}
\label{fig:FERET-samples}
\end{figure}

The morphed images are created using FaceFusion ~\cite{FaceFusi30:online} and UBO Morpher ~\cite{ferrara2019decoupling}. More information about the images used is provided in the table \ref{table:number-of-images}. Figures \ref{fig:FERET-samples} and \ref{fig:FRGC-samples} show sample images from FERET and FRGCv2 datasets and their morphs created using FaceFusion and UBO Morpher.

\begin{table}[htbp]
\centering
\caption{Number of images in the datasets}
\begin{tabular}{c c c c}
\hline
Dataset & Bonafide & FaceFusion & UBO Morpher \\ \hline
FERET & 622 & 529 & 529 \\ 
FRGCv2 & 1440 & 964 & 964 \\
\end{tabular}
\label{table:number-of-images}
\end{table}

\subsection{Evaluation metrics}
The proposed solution is tested based on the metrics established by the ISO/IEC 30107-3 for performance assessment of presentation attack detection mechanisms i.e. attack presentation classification error rate (APCER) and bona fide presentation classification error rate (BPCER) ~\cite{biometrics2016iso}. APCER is a measure of falsely accepting the morphed images as bonafide images.  Whereas, the BPCER is a measure of falsely rejecting the bonafide images by classifying them as morphed images. Here, the detection equal error rate (D-EER) point (where APCER=BPCER) is reported, along with the BPCER10 (where APCER = 10\%) and BPCER20 (where APCER = 5\%), as described in ~\cite{debiasi2018prnu}. 

\subsection{Experiment setup and results}
The experiments were carried out under different settings and with various datasets to fine-tune and evaluate the reliability of the proposed method. The final results are hereby reported in this section.

\subsubsection{Effect of increasing contrast}
\label{section:contrast-results}
\begin{figure}
    \centering
    \includegraphics[width=0.5\textwidth]{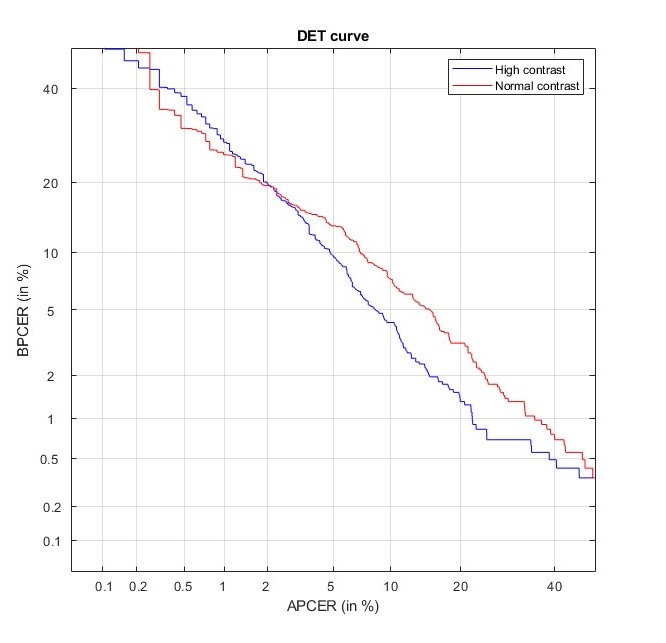}
    \caption{DET curves by varying contrast}
    \label{fig:det-curve-contrast}
\end{figure}

For the preprocessing step, we experimented with increasing the contrast of the cropped eyebrow region. Since image contrast can be used to enhance the differentiation in the textures present in the image, the idea is that this can further increase the frequency content of the bonafide images as compared to the morphed images. Figure \ref{fig:det-curve-contrast} shows the DET curves by varying the contrast on the FRGC images. The ISO metrics are presented in table \ref{table:effect-of-contrast} where it can be seen that the assumption is correct showing that increasing contrast helps improve the efficiency of the system.

\begin{table}[htb]
\centering
\caption{Effect of enhancing contrast on morph detection}
\begin{tabular}{c c c c}
\hline
Image Enhancement & D-EER (\%) & BPCER10 (\%) & BPCER20 (\%) \\ \hline
No & 8.5 & 7.3 & 13.9 \\ 
Yes (high contrast) & 6.5 & 4.2 & 9.6 \\ 
\end{tabular}
\label{table:effect-of-contrast}
\end{table}

\subsubsection{Effect of cropping low-frequency content}
Since the eyebrows are associated with the presence of high-frequency content, we experimented with cropping the low-frequency coefficients in the frequency spectrum. The experiments were conducted by ignoring 0\%, 5\%, and 10\% of the low-frequency region from the calculations. The results are presented in table \ref{table:frequency-cropping} which show that cropping the low-frequency region slightly reduces the performance of morphed image detection. Since the results are against the proposed idea, this step was not incorporated in the final proposed algorithm.

\begin{table}[htb]
\centering
\caption{Effect of ignoring low-frequency component}
\begin{tabular}{c c c c}
\hline
Crop (\%) & D-EER (\%) & BPCER10 (\%) & BPCER20 (\%) \\ \hline
0 & 6.5 & 4.2 & 9.6 \\  
5 & 6.6 & 3.8 & 9.9 \\ 
10 & 6.7 & 4.1 & 9.8 \\ 
\end{tabular}
\label{table:frequency-cropping}
\end{table}


\subsubsection{Final results on different datasets}
Table \ref{table:performance-with-different-datasets} shows the results by applying the proposed scheme on the two datasets separately and then combining them. It can be seen that the proposed scheme gives a much better result with the FRGC dataset than with the FERET dataset. D-EER of 6.5\% obtained with the FRGCv2 dataset is considerably lower than the D-EER of 22.2\% on the FERET dataset.  

\begin{table}[htb]
\centering
\caption{Detection performance with different datasets}
\begin{tabular}{c c c c}
\hline
Dataset & D-EER (\%) & BPCER10 (\%) & BPCER20 (\%) \\ \hline
FRGCv2 & 6.5 & 4.2 & 9.6 \\ 
FERET & 22.2 & 38.2 & 51.7 \\
Combined & 14.2 & 17.02 & 23.2 \\ 
\end{tabular}
\label{table:performance-with-different-datasets}
\end{table}

\begin{figure}[ht]
    \centering
    \includegraphics[width=0.5\textwidth, height=5cm]{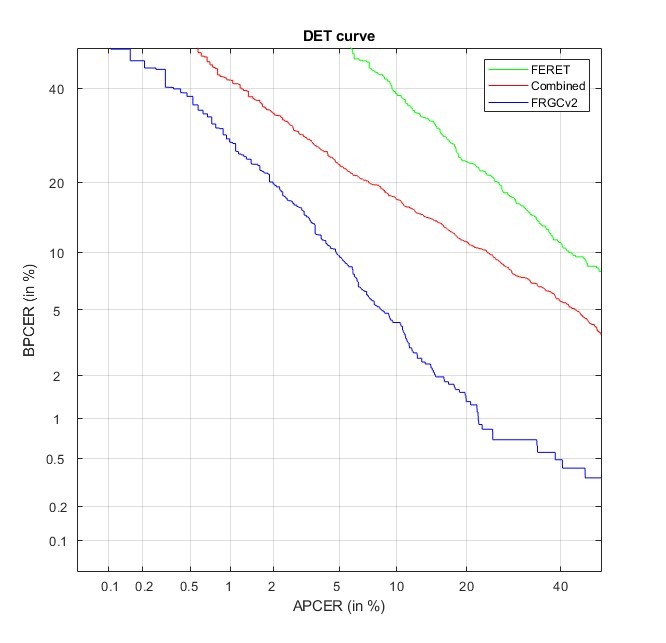}
    \caption{DET curves obtained by applying the proposed scheme on different datasets}
    \label{fig:det-curves-different-datasets}
\end{figure}

On inspecting the images, it was found that the FERET images have lower quality compared to the FRGC dataset. This can also be attributed to the fact that the FERET images are relatively old (from 2011) compared to the FRGCv2 images (from 2014). Hence, due to the lower resolution of the images in the FERET dataset, it is harder to differentiate between a bonafide image and a morphed image. This claim is supported by the sizes of files from both datasets as shown in the table \ref{table:filesizes-in-datasets}. \red{These results are also supported by the findings in  ~\cite{scherhag2020face}, where all MAD algorithms based on texture descriptors gave better results for the FRGC dataset compared to FERET.}

\begin{table}[htb]
\centering
\caption{Bonafide image sizes among different datasets (in Kb)}
\begin{tabular}{c c c c}
\hline
Dataset  & Min & Max & Average  \\ \hline
FERET & 704.4 & 1218.6 & 924.7 \\ 
FRGCv2 & 344.4 & 1065.8 & 722.2 \\ 
\end{tabular}
\label{table:filesizes-in-datasets}
\end{table}


\subsubsection{Comparison with previous work}



In this section, the results will be presented by comparing them with the previous S-MAD techniques. For comparing the results with ~\cite{ndeh2021morphed}, experiments are performed on the FRGC dataset by dividing the dataset into training and testing subsets. The resulting error rates are presented in table \ref{table:frgc-error-rates}. These also include ACER which is the average classification error rate ~\cite{ndeh2021morphed}. These error rates are higher than the ones obtained in ~\cite{ndeh2021morphed} showing that the proposed scheme does not improve the detection performance. However, if we consider the overall performance on different datasets, the proposed scheme gives a lower D-EER of 22.2\% in comparison with ACER of 38.24\% in the other paper.

\begin{table}[htbp]
\centering
\caption{BPCER, APCER, and ACER values with FRGC dataset}
\begin{tabular}{c|c| c| c|c|c|c}
\hline
Subjects & Total & Rightly classified & Wrongly classified & APCER & BPCER & ACER \\ \hline
Bonafide & 720 & 680 & 40 & 5.5 & 4.9 & 5.2 \\
Morphs & 964 & 916 & 48 &  &  & \\ 
\end{tabular}
\label{table:frgc-error-rates}
\end{table}

\section{Conclusion and Discussion}
\label{section:conclusion}
The results indicate the effectiveness of the proposed method in detecting morphed images. The frequency spectrum analysis of the eyebrow region proves to be a promising approach for the detection of morphed images in light of the results. In addition to the S-MAD scenario, the proposed method can also be applied in a D-MAD system where a reduction in error rate is expected. Even though the detection capabilities were found to be dependent on the choice of dataset, these results were expected given the quality of images varied between the datasets, as explained in the results section. However, the detection capabilities were found to be robust against two different kinds of morphing techniques.

There are some limitations of using this approach that need to be investigated further. 1) People with certain diseases can not have eyebrows. Our datasets do not contain any such cases and hence the result of the segmentation method and frequency analysis in these cases needs can be studied for improving this method. 2) The morphed images used in the experiments are created from automated morphing algorithms. For an attacker to conduct a successful attack, a manually generated image is sufficient. So, further analysis can be made to study the possibility of altering a morphed image to bypass the proposed detection scheme.

%
%
%
 \bibliographystyle{splncs04}
 \bibliography{references}
\end{document}